\documentclass{SciPost}

\hypersetup{
    colorlinks,
    linkcolor={red!50!black},
    citecolor={blue!50!black},
    urlcolor={blue!80!black}
}
\usepackage{soul}
\usepackage{tcolorbox}
\usepackage[dvipsnames]{xcolor}
\usepackage{braket}
\usepackage{tikz}
\usepackage{subcaption}
\usetikzlibrary{arrows.meta,decorations.markings}
\usepackage[bitstream-charter]{mathdesign}
\usetikzlibrary{calc}
\usetikzlibrary{shapes.geometric, arrows.meta, positioning}
\tikzstyle{start} = [rectangle, rounded corners, minimum width=3cm, minimum height=1cm,text centered, draw=black, fill=red!30]
\tikzstyle{stop} = [rectangle, rounded corners, minimum width=3cm, minimum height=1cm,text centered, draw=black, fill=green!30]
\tikzstyle{process} = [rectangle, minimum width=3cm, minimum height=1cm, text centered, draw=black, fill=blue!20]
\tikzstyle{decision} = [diamond, aspect=2, minimum width=3cm, minimum height=1cm, text centered, draw=black, fill=orange!30]
\tikzstyle{arrow} = [thick, ->, >=stealth]

\DeclareSymbolFont{usualmathcal}{OMS}{cmsy}{m}{n}
\DeclareSymbolFontAlphabet{\mathcal}{usualmathcal}

\fancypagestyle{SPstyle}{
\fancyhf{}
\lhead{\colorbox{scipostblue}{\bf \color{white} ~SciPost Physics }}
\rhead{{\bf \color{scipostdeepblue} ~Submission }}

\fancyfoot[C]{\textbf{\thepage}}
}

\newcommand{\ham}{\hat{\cal H}}

\begin{document}

\pagestyle{SPstyle}

\begin{center}{\Large \textbf{\color{scipostdeepblue}{
Quantitative approach for the Dicke-Ising chain with an effective self-consistent matter Hamiltonian 
}}}\end{center}

\begin{center}\textbf{
Jonas Leibig\textsuperscript{$\star$},
Max Hörmann,
Anja Langheld,
Andreas Schellenberger and 
\mbox{Kai Phillip Schmidt\textsuperscript{$\dagger$}} 
}\end{center}

\begin{center}
Department of Physics, Staudtstraße 7, Friedrich-Alexander-Universität Erlangen-Nürnberg, Germany
\\[\baselineskip]
$\star$ \href{mailto:email1}{\small jonas.leibig@fau.de}\,,\quad
$\dagger$ \href{mailto:email2}{\small kai.phillip.schmidt@fau.de}
\end{center}

\begin{abstract}
In the thermodynamic limit, the Dicke-Ising chain maps exactly onto an effective self-consistent matter Hamiltonian with the photon field acting solely as a self-consistent effective field. As a consequence, no quantum correlations between photons and spins are needed to understand the quantum phase diagram. 
This enables us to determine the quantum phase diagram in the thermodynamic limit using numerical linked-cluster expansions combined with density matrix renormalization group calculations (NLCE+DMRG) to solve the resulting self-consistent matter Hamiltonian. This includes magnetically ordered phases with significantly improved accuracy compared to previous estimates. 
For ferromagnetic Ising couplings, we refine the location of the multicritical point governing the change in the order of the superradiant phase transition, reaching a relative accuracy of $10^{-4}$. 
For antiferromagnetic Ising couplings, we confirm the existence of the narrow 
antiferromagnetic superradiant phase in the thermodynamic limit. 
The effective matter Hamiltonian framework identifies the antiferromagnetic superradiant phase as the many-body ground 
state of an antiferromagnetic transverse-field Ising model with longitudinal field. 
This phase emerges through continuous Dicke-type polariton condensation from the antiferromagnetic normal phase, followed by a first-order transition to the paramagnetic 
superradiant phase.
Thus, NLCE+DMRG provides a precise determination of the Dicke-Ising phase diagram in one dimension by solving the self-consistent 
effective matter Hamiltonian.
\end{abstract}

\vspace{\baselineskip}

\noindent\textcolor{white!90!black}{%
\fbox{\parbox{0.975\linewidth}{%
\textcolor{white!40!black}{\begin{tabular}{lr}%
  \begin{minipage}{0.6\textwidth}%
    {\small Copyright attribution to authors. \newline
    This work is a submission to SciPost Physics. \newline
    License information to appear upon publication. \newline
    Publication information to appear upon publication.}
  \end{minipage} & \begin{minipage}{0.4\textwidth}
    {\small Received Date \newline Accepted Date \newline Published Date}%
  \end{minipage}
\end{tabular}}
}}
}


\vspace{10pt}
\noindent\rule{\textwidth}{1pt}
\tableofcontents
\noindent\rule{\textwidth}{1pt}
\vspace{10pt}


\section{Introduction}

The Dicke model~\cite{dicke1954coherence}, which describes $N$ two-level systems collectively coupled to a single bosonic mode, is a paradigmatic model of collective light-matter interactions.  
In the thermodynamic limit $N \to \infty$, Hepp and Lieb~\cite{HeppLieb1973} proved that the Dicke model exhibits a quantum phase transition from a normal phase to a superradiant phase, characterized by macroscopic photon occupation and spontaneous $\mathbb{Z}_2$ symmetry breaking.  
An alternative derivation of this transition was later provided by Wang and Hioe~\cite{wang1973}, who evaluated the exact free energy using Glauber coherent states and demonstrated the phase transition in an elementary and transparent manner. Their results coincide precisely with those of Hepp and Lieb and can be generalized to multiple radiation modes.  

The cavity-QED implementation proposed by Dimer \emph{et al.}~\cite{dimer2007}, who showed that cavity-mediated Raman transitions can realize an effective Dicke Hamiltonian in the critical regime, directly inspired the subsequent experimental observations of the normal-to-superradiant transition with ultracold atoms in optical resonators~\cite{baumann2010, baumann2011}.  
Related signatures of collective light-matter coupling have also been explored in circuit-QED platforms, where ensembles of superconducting qubits interact collectively with a microwave cavity mode~\cite{fink2009}. 

 On the theoretical side, the Dicke model can also be mapped to an all-to-all interacting spin system. This connection to the Lipkin-Meshkov-Glick model was pioneered by Reslen \emph{et al.}~\cite{reslen2004direct} using cumulant expansion methods. Subsequently, displacement transformation techniques~\cite{rohn} and path-integral formulations with saddle-point analysis~\cite{roman2025bound,roman2025linear} rigorously established that in the thermodynamic limit $N \to \infty$ Dicke-type models map exactly to effective matter Hamiltonians. Related approaches based on nonlinear susceptibilities~\cite{lenk2022collective} further confirmed this decoupling. A crucial consequence of this exact mapping is that all extensive observables and thus the phase diagram can be obtained from an effective matter Hamiltonian.
In this formulation, the photon field appears only as a self-consistent parameter modifying the matter sector. In the specific single-mode setting considered here, this reframes the thermodynamic limit as a controlled matter problem where light acts mainly as a tunable parameter rather than a quantum degree of freedom that qualitatively influences the phase diagram. 

It is natural to extend the Dicke model to include direct spin-spin (e.g. Ising-type) interactions. In many platforms, such as superconducting qubit arrays with Josephson coupling or atomic systems where short-range interactions are relevant, nearest-neighbor couplings between the two-level systems arise and can compete with the collective light-matter coupling. This naturally leads to the Dicke-Ising model, where collective light-matter coupling is combined with local Ising interactions. Digital-analog quantum simulators for the realization of the Dicke-Ising model based on an ensemble of interacting qubits coupled to a single-mode photonic resonator have also been proposed \cite{Shapiro2025}.  

Zhang \emph{et al.}~\cite{Zhang2014} carried out a mean-field analysis of a circuit-QED implementation of the Dicke-Ising model, realized by a one-dimensional array of superconducting qubits with antiferromagnetic nearest-neighbor coupling. Their study revealed a competition between the collective spin-photon interaction and the antiferromagnetic Ising term and predicted four distinct quantum phases, leading to an antiferromagnetic superradiant phase in which antiferromagnetic and superradiant orders coexist. These predictions motivate a more controlled understanding of the Dicke-Ising model beyond mean-field theory. 
Schellenberger and Schmidt~\cite{Schellenberger_2024} developed such an approach by mapping correlated light-matter models onto an exactly solvable Dicke model in the non-superradiant phases. Applied to the Dicke-Ising system, this mapping shows that the Dicke-like excitations and the mean-field phase boundary coincide by solving the low-energy properties in the non-superradiant regime.

Rohn \emph{et al.}~\cite{rohn} studied the special limit of a vanishing longitudinal magnetic field, the so-called quantized transverse-field Ising model (QTFIM). In this limit, the effective matter Hamiltonian maps, in one dimension, to the exactly solvable transverse-field Ising model and they predicted a first-order quantum phase transition from a magnetically ordered normal phase to a paramagnetic superradiant phase.  
Further evidence for first-order behavior in the QTFIM was provided by Gammelmark \emph{et al.}~\cite{gammelmark2011}, who analyzed finite spin chains with collective cavity coupling, and by Lee and Johnson~\cite{lee2004}, who demonstrated that multiqubit cavity systems can generically exhibit first-order superradiant transitions.  
The first-order character in the QTFIM was subsequently confirmed through quantum Monte Carlo simulations using the wormhole algorithm~\cite{langheld} and through the analysis of bound polariton states~\cite{roman2025bound}.

In addition to these analytical insights, large-scale numerical simulations have recently become feasible. Quantum Monte Carlo (QMC) studies by Langheld \emph{et al.} mapped out the phase diagram of the Dicke-Ising model on the chain and the square lattice~\cite{langheld}. Their results confirmed the emergence of an antiferromagnetic superradiant phase and identified the nature of the transition lines, including a multicritical point separating continuous and first-order transitions~\cite{langheld}. Most recently, Mendonça \emph{et al.}~\cite{mendoncca2025} used density matrix renormalization group (DMRG) simulations on the Dicke-Ising chain. Their study reported that no phase with coexisting antiferromagnetic and superradiant order is observed and that the normal-to-superradiant transition remains essentially unchanged from the pure Dicke case, contradicting and ignoring previous works \cite{rohn, roman2025linear, roman2025bound, langheld, Zhang2014} including quantitative QMC simulations as outlined in the comment Ref.~\cite{comment2025}. 

In this work we investigate the Dicke-Ising chain via a mapping to an effective self-consistent matter Hamiltonian. While this mapping is exact in the thermodynamic limit, the effective Hamiltonian becomes non-trivial to solve whenever the system develops a finite photon field, since the resulting self-consistency condition couples back to matter correlations; in the normal phase, where the photon field vanishes, this complication is absent.
We solve the self-consistent matter Hamiltonian by numerical linked-cluster expansions \cite{rigol2006,rigol2007a,rigol2007b} combined with density-matrix renormalization group \cite{White1992, White1993} (NLCE+DMRG), a framework that operates directly in the thermodynamic limit where the light-matter decoupling becomes exact. Applying it to the Dicke-Ising chain, we determine the ferromagnetic multicritical point with an accuracy of \(10^{-4}\). For the antiferromagnetic case, we confirm the existence and establish the microscopic origin of the narrow antiferromagnetic-superradiant phase.

The remainder of this paper is organized as follows. In \autoref{sec:DImodel}, we introduce the Dicke-Ising model and derive an effective matter Hamiltonian in the thermodynamic limit in \autoref{sec:effective_model}. Our NLCE+DMRG approach is described in detail in \autoref{sec:method}. The results are divided into two subsections, with \autoref{sec:ferro} discussing the multicritical point for ferromagnetic Ising interactions, while \autoref{sec:af} demonstrates the existence of an antiferromagnetic superradiant phase in one dimension for antiferromagnetic Ising interactions. Finally, we summarize our findings and provide an outlook in \autoref{sec:conclusion}.

\section{Model}
\subsection{Dicke-Ising model}
\label{sec:DImodel}

The Dicke-Ising model (DIM) combines the paradigmatic Dicke model \cite{dicke1954coherence} of quantum optics with nearest-neighbor Ising interactions. It describes a system of \( N \) spin-\(1/2\) degrees of freedom in a level-splitting field that are coupled to a single-mode bosonic field and interact via a nearest-neighbor Ising term. The Hamiltonian reads
\begin{equation}
    \hat{H}_{\text{DIM}} = 2\varepsilon \hat{S}_z + \frac{g}{\sqrt{N}} (\hat{a}^\dagger + \hat{a}) \hat{S}_x + \omega_c \hat{a}^\dagger \hat{a} + J \sum_{\langle i, j \rangle} \hat{\sigma}_i^z \hat{\sigma}_j^z\, ,
    \label{eq:DIM}
\end{equation}
where \( 2\varepsilon \) denotes the energy splitting of the spin states (or equivalently, the strength of a longitudinal field), and \( \hat{a}^\dagger \) (\( \hat{a} \)) are bosonic creation (annihilation) operators for the photon mode of frequency \( \omega_c \). The spin operators are defined collectively as \( \hat{S}_\alpha = \sum_i \hat{\sigma}_i^\alpha/2 \) for \( \alpha \in \{x, y, z\} \,\), and satisfy the SU(2) algebra: \( [\hat{S}_\alpha, \hat{S}_\beta] = \mathrm{i} \varepsilon_{\alpha\beta\gamma} \hat{S}_\gamma \,\). The spin-photon coupling strength is given by \( g \), and \( J \) denotes the Ising interaction strength between neighboring spins. The interaction is ferromagnetic for \( J < 0 \) and antiferromagnetic for \( J > 0 \,\).

In the absence of the longitudinal field (\( \varepsilon = 0 \)), the DIM possesses a \( \mathbb{Z}_2 \times \mathbb{Z}_2 \) symmetry: one \( \mathbb{Z}_2 \) symmetry originates from the Dicke model as a combined parity transformation \( (\hat{a}, \hat{S}_x) \to (-\hat{a}, -\hat{S}_x) \), and the other from the Ising model via the spin-flip operation \( \hat{\sigma}^z \to -\hat{\sigma}^z \). For finite~$\varepsilon$, while the spin-flip symmetry is broken explicitly by the magnetic field, the Hamiltonian keeps the parity symmetry associated with the Dicke model and the discrete lattice translation symmetry. 
The essential difference between ferromagnetic ($J<0$) and antiferromagnetic ($J>0$) interactions lies not in the symmetry of the Hamiltonian, but in which of these symmetries can be spontaneously broken in the ground state.
A ferromagnetically ordered state preserves the discrete lattice translation symmetry and may only break parity such that phases with simultaneous breaking of two symmetries cannot occur. 
In contrast, for antiferromagnetic couplings, the discrete lattice translation symmetry can be spontaneously broken, leading to a two-sublattice structure characteristic of antiferromagnetic order. This allows both parity and lattice translation symmetry to be broken simultaneously, enabling the antiferromagnetic superradiant (AS) phase.

\begin{figure}[t]
    \centering
    \includegraphics{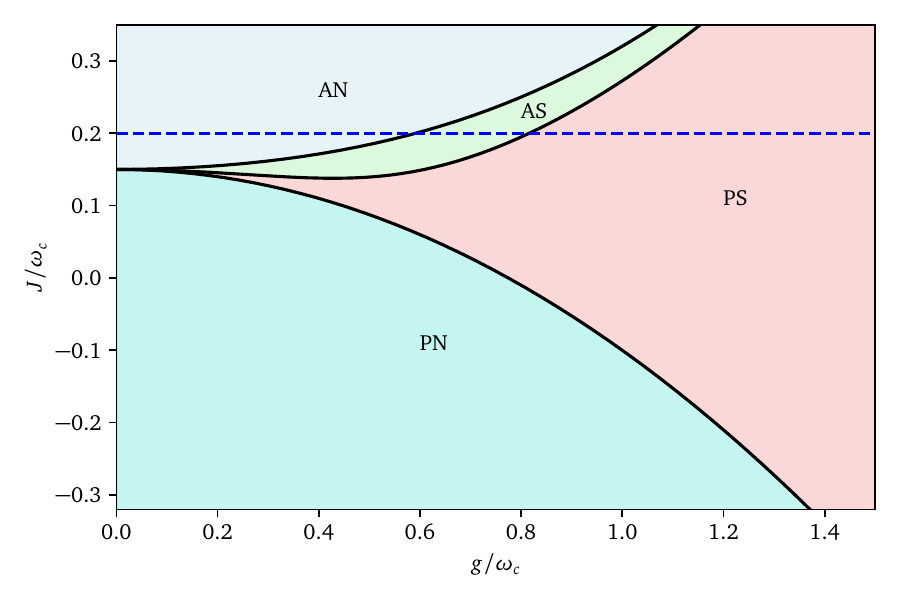}
    \caption{Mean-field phase diagram of the Dicke-Ising model at fixed longitudinal field \( \varepsilon/\omega_c = 0.3 \), plotted as a function of the spin-photon coupling \( g/\omega_c \) and the Ising interaction strength \( J/\omega_c \) \cite{Zhang2014}. Four phases are identified: paramagnetic normal (PN), paramagnetic superradiant (PS), antiferromagnetic normal (AN), and antiferromagnetic superradiant (AS). The AS phase breaks two discrete symmetries and emerges only for antiferromagnetic couplings. Additonally, the sweep in \(g\) from \autoref{subsec:af_numerical} is visualized as a dashed blue line.}
    \label{fig:mf_phase_gJ}
\end{figure}

Zhang~\emph{et al.}~\cite{Zhang2014} performed a mean-field study based on a variational product state consisting of a coherent photonic state and a classical spin state with a two-site spin unit cell.
We present their mean-field phase diagram \cite{Zhang2014} in Fig.~\ref{fig:mf_phase_gJ}. The phase diagram is visualized in terms of the spin-photon coupling \( g/\omega_c \) and the Ising interaction strength \( J/\omega_c \), for a fixed longitudinal field \( \varepsilon/\omega_c = 0.3 \). This allows us to compare the ferromagnetic (\( J < 0 \)) and antiferromagnetic (\( J > 0 \)) regimes within a single graphic. As shown in Fig.~\ref{fig:mf_phase_gJ}, the mean-field analysis predicts four distinct phases: the paramagnetic normal (PN) phase, the paramagnetic superradiant (PS) phase, the antiferromagnetic normal (AN) phase, and the antiferromagnetic superradiant (AS) phase. The AS phase is particularly noteworthy as it features simultaneous spin and photonic order, thereby breaking both the discrete lattice translation and the parity symmetry. While mean-field theory predicts continuous phase transitions for $\varepsilon\neq0$ ~\cite{Zhang2014}, it is important to note that at the special point $\varepsilon = 0$ the mean-field transition becomes first order. 

Large-scale quantum Monte Carlo simulations confirm that the first-order character persists for small longitudinal fields~\cite{langheld} in one and two dimensions, in disagreement with the immediately continuous transition predicted by mean-field theory. The crossover from first- to second-order behavior was estimated to occur near $J \approx - \varepsilon$~\cite{langheld}. For antiferromagnetic Ising interactions ($J>0$), the AS phase remains present but is extremely narrow in one dimension~\cite{langheld}. In addition, recent DMRG work by Mendonça \emph{et al.}~\cite{mendoncca2025} reported the absence of the AS phase and concluded that the Dicke-Ising chain behaves similarly to the pure Dicke model. As discussed in detail in Ref.~\cite{comment2025}, these claims are inconsistent with previously established analytical and numerical evidence~\cite{rohn,roman2025linear,roman2025bound,langheld,Zhang2014}.

Finally, let us discuss the influence of a diamagnetic \(A^2\)-term, which commonly arises in realizations of Dicke systems \cite{minimalcoupling1,minimalcoupling2, minimalcoupling3} and can be investigated within the framework of the Hamiltonian in Eq.~\eqref{eq:DIM} \cite{langheld}. The \(A^2\)-term takes the form
\begin{equation}
\hat{H}_{A^2} = D\left( \hat{a}^\dagger + \hat{a}\right)^2,
\end{equation}
where this contribution naturally emerges from the microscopic derivation of the Dicke-Ising model using the minimal coupling substitution \cite{minimalcoupling1,minimalcoupling2,minimalcoupling3}. In the limit of the Dicke model (\(J=0\)), this diamagnetic term acts to suppress the formation of a superradiant phase \cite{minimalcoupling1}.

In the Dicke-spin models like Hamiltonian \eqref{eq:DIM}, the presence of the \(A^2\)-term can be effectively incorporated into the photon energy via a Bogoliubov transformation \cite{minimalcoupling2}. This transformation renormalizes both the photon frequency and the spin-photon coupling according to
\begin{align}
\omega' &= \sqrt{\omega_c^2 + 4\omega_c D}\,,& \
g' &= \sqrt{\frac{\omega_c}{\omega'}}\,g.
\label{eq:renormalized_params}
\end{align}
The parameter \(D\) is not independent, as the Thomas-Reiche-Kuhn (TRK) sum rule imposes a lower bound given by \(D \geq g^2/2\varepsilon\) depending on the concrete experimental setup \cite{minimalcoupling2, superrandiantMarquardt}. The effective couplings do not qualitatively alter the phase structure. In particular, all four phases remain accessible for antiferromagnetic Ising interactions. However, for ferromagnetic Ising interactions the superradiant phase is not realizable.

\subsection{Mapping to a self-consistent effective model}
\label{sec:effective_model}
We will describe the Dicke-Ising model using an effective spin Hamiltonian in which the cavity's influence is captured by a self-consistent transverse field. This mapping is exact in the thermodynamic limit and will enable efficient numerical access to ground-state properties.
We start from the Dicke-Ising Hamiltonian \eqref{eq:DIM} and apply a displacement transformation~\cite{rohn} with the unitary operator
\begin{equation}
\hat{D}(\gamma) = \exp\!\left[\gamma\, \hat{S}_x (\hat{a}^\dagger - \hat{a})\right], \quad \text{with} \quad \gamma = \frac{g}{\omega_c \sqrt{N}}\, .
\end{equation}
The transformed Hamiltonian becomes
\begin{equation}
\hat{H}'_{\text{DIM}} = \hat{D}(\gamma)\, \hat{H}_{\text{DIM}}\, \hat{D}^\dagger(\gamma)\, .
\end{equation}
Expanding $\hat{H}'_{\text{DIM}}$ in the small parameter $\gamma$ yields
\begin{equation}
\hat{H}'_{\text{DIM}} \;=\; \omega_c\, \hat{a}^\dagger \hat{a} \;-\; \frac{g^2}{\omega_c\, N}\, \hat{S}_x^2 \;+\; 2\varepsilon\, \hat{S}_z \;+\; J \sum_{\langle i, j \rangle} \hat{\sigma}_i^z \hat{\sigma}_j^z \;+\; \hat{R}\left(\gamma^2\right)\, ,
\end{equation}
where the remainder term $\hat{R}$ scales sub-extensively, as outlined in the supplement material of Ref.~\cite{langheld}. More precisely, the remainder term contributes lower than $O(N)$ in the extensive energy. Therefore, $\hat{R}$ does not affect the energy per particle for $N\rightarrow\infty$. This makes the approximation exact in the thermodynamic limit for extensive ground-state observables, which determine the zero-temperature phase diagram. 

An advantage of applying the displacement transformation before the mean-field decoupling is that it provides a systematic hierarchy of $1/N$ corrections visible in the Hamiltonian picture through the remainder term $\hat{R}$. The path-integral formulation with saddle-point expansion~\cite{roman2025linear,roman2025bound} provides complementary insight into the $1/N$ correction structure, though exact treatment of higher-order corrections remains an open challenge in both frameworks.

After the displacement transformation, the light-matter coupling manifests as the nonlinear $\hat{S}_x^2/N$ term, representing an induced all-to-all interaction among the spins. The subsequent treatment of this $\hat{S}_x^2/N$ term allows mean-field decoupling. While Ref.~\cite{rohn} introduced the displacement transformation, the self-consistent mean-field treatment was not correctly implemented. The correct treatment was established in Refs.~\cite{roman2025linear,roman2025bound,langheld}, and applied to determine the exact transition point for the Dicke-Ising chain with $\varepsilon=0$ in Refs.~\cite{roman2025bound,langheld}. 

Next, we apply this mean-field decoupling to the nonlinear term $\hat{S}_x^2$. Since this term couples all spins uniformly and scales as $1/N$, fluctuations become negligible in the thermodynamic limit and the mean-field treatment becomes exact~\cite{roman2025linear, roman2025bound, langheld}.
We write
\begin{equation}
\hat{S}_x^2 \approx -\,M_x^2 + 2 M_x \hat{S}_x\, ,
\end{equation}
where $M_x = \langle \hat{S}_x \rangle$ denotes the total transverse magnetization in $x$-direction. Applying this decoupling, the resulting self-consistent Hamiltonian in the subspace of a vanishing photon number reads
\begin{equation}
\hat{H}'_{\text{eff}}(m_x) = \frac{g^2 N}{\omega_c} m_x^2 - \frac{g^2}{\omega_c} m_x \sum_i \hat{\sigma}_i^x + \varepsilon \sum_i \hat{\sigma}_i^z + J \sum_{\langle i, j \rangle} \hat{\sigma}_i^z \hat{\sigma}_j^z\,,
\label{eq:Heff}
\end{equation}
with $m_x = M_x/N$.
The effect of the cavity mode thus appears as an effective transverse field $h_x \equiv \frac{g^2}{\omega_c} m_x$ 
, which must be determined self-consistently by requiring $m_x \equiv \langle \sum_i \hat{\sigma}_i^x \rangle/N$ in the ground state. The first term in the effective Hamiltonian is an energy shift that is constant for a given spin eigenstate, i.e., it depends on the spin configuration but not on the photon dynamics.

It is instructive to trace the fate of the $\mathbb{Z}_2$ symmetry stemming from the Dicke model through the 
mapping. This symmetry corresponds to the 
combined parity transformation $(\hat{a}, \hat{S}^x) \to (-\hat{a}, -\hat{S}^x)$. 
After the displacement transformation, the photon sector resides in the vacuum 
state, which is invariant under the photon parity transformation $\hat{a} \to -\hat{a}$. 
The $\mathbb{Z}_2$ symmetry therefore manifests purely in the spin sector as 
$\hat{S}^x \to -\hat{S}^x$. The mean-field decoupling of $\hat{S}_x^2$ then 
explicitly breaks this operator symmetry, since $m_x$ enters as a fixed parameter. 
However, the symmetry is not lost but transferred to the corresponding energy functional $E(m_x)\equiv \langle \hat{H}'_{\text{eff}} \rangle$, which 
satisfies $E(m_x) = E(-m_x)$. Spontaneous symmetry breaking in the superradiant 
phases corresponds to the system selecting one of two equivalent minima at $\pm m_x$.

To obtain the ground-state properties, we solve the Hamiltonian for a given trial 
value of $m_x$, compute the resulting magnetization $\langle \hat{S}^x \rangle / N$ and set it as the next trial value for $m_x$, which we iterate until self-consistency is achieved. Equivalently, this procedure 
minimizes $E(m_x)$ with respect to $m_x$, as shown below.

We note that the self-consistency condition $m_x = \langle \hat{S}_x \rangle/N$ and the condition $\frac{dE}{dm_x} = 0$ are equivalent via the Hellmann-Feynman theorem.  Specifically, 
\begin{equation}
\frac{dE}{dm_x} = 2\frac{g^2}{\omega_c} N m_x + \frac{dE_{\text{matter}}}{dm_x},
\end{equation}
with 
\begin{equation}
    E_{\text{matter}}= \left\langle - \frac{g^2}{\omega_c} m_x \sum_i \hat{\sigma}_i^x + \varepsilon \sum_i \hat{\sigma}_i^z + J \sum_{\langle i, j \rangle} \hat{\sigma}_i^z \hat{\sigma}_j^z\, \right\rangle\;.
\end{equation}
The Hellmann-Feynman theorem gives
\begin{equation}
    \frac{dE_{\text{matter}}}{dm_x} = -\frac{g^2}{\omega_c} \left\langle \sum_i \hat{\sigma}_i^x \right\rangle = -2\frac{g^2}{\omega_c} N \langle \hat{S}_x \rangle/N\,.
\end{equation}
Therefore, $\frac{dE}{dm_x} = 0$ if and only if $m_x = \left\langle \hat{S}_x \right\rangle/N$. Self-consistent solutions correspond to stationary points of the total energy functional $E(m_x)$. The ground state is obtained by selecting the solution with lowest energy. This establishes that the self-consistency condition $m_x = \langle \hat{S}^x \rangle / N$ and the condition $\mathrm{d}E/\mathrm{d}m_x = 0$ are equivalent rather than independent requirements. 

The effective model \eqref{eq:Heff} captures all relevant ordering phenomena of the Dicke-Ising model in the thermodynamic limit. Photon observables such as the photon density can be obtained via the displacement transformation from the self-consistent transverse magnetization. The effective model forms the basis of our NLCE+DMRG approach for light-matter quantum systems and the obtained quantum phase diagram for the Dicke-Ising chain discussed in the remainder of this work.

\section{Method: NLCE+DMRG} 

\label{sec:method}
In this section, we introduce our NLCE+DMRG approach, which is used to numerically calculate the ground-state phase diagram of the self-consistent matter Hamiltonian Eq.~\eqref{eq:Heff} for the Dicke-Ising chain. Our approach is based on numerical linked cluster expansions (NLCE) to directly access the thermodynamic limit. As a cluster solver for the NLCE we replace the commonly used exact diagonalization by density matrix renormalization group (DMRG) calculations. This procedure was already used in Ref.~\cite{stoudemire14} to calculate the corner contribution to the entanglement entropy of strongly interacting O(2) quantum critical systems in 2+1 dimensions.

The numerical convergence of NLCE+DMRG is controlled solely by the finite correlation length of the self-consistent effective matter Hamiltonian \eqref{eq:Heff}. 
This is different to finite-size effects of the light-matter system, e.g., the calculations with QMC in Ref.~\cite{langheld} or with DMRG in Ref.~\cite{Sur2025}, where the extensive photon contribution introduces size-dependent corrections that vanish only in the limit $N \to \infty$. 
This fundamental difference can be illustrated by the pure Dicke model setting $J = 0$ in \autoref{eq:DIM}: the effective matter Hamiltonian \eqref{eq:Heff} reduces to a $2 \times 2$ matrix problem that is solved exactly by NLCE+DMRG with single-site clusters, requiring no finite-size extrapolation. In contrast, the direct treatment of the Dicke model on finite clusters exhibit finite-size corrections, requiring extrapolation to the thermodynamic limit. However, the QMC algorithm developed in Ref.~\cite{langheld} for the Dicke-Ising model provides numerically precise results for large finite systems up to several thousands of spins, which our thermodynamic-limit NLCE+DMRG approach cannot access.

The main challenge for the NLCE lies in satisfying the self-consistency condition of the Hamiltonian in Eq.~\eqref{eq:Heff}. Antiferromagnetic Ising interactions further complicate this, as the magnetically ordered phase of the one-dimensional chain must be described in terms of a two-site unit cell with alternating spin orientations. Within the NLCE framework, one of the two symmetry-equivalent antiferromagnetic configurations (starting with spin up or spin down) must be chosen to ensure a consistent ordering across clusters. To this end we will introduce boundary fields on the clusters, reflecting the coupling to the ordered environment. These boundary fields have to be determined self-consistently, thereby introducing a second self-consistency condition in the Hamiltonian.

In the following we briefly describe NLCE in one dimension, give details about the used DMRG implementation, and explain how to solve the self-consistency equations within the NLCE+DMRG approach.

\subsection{NLCE in 1D}

NLCE provides a framework to compute extensive quantities directly in the thermodynamic limit by systematically summing up contributions from finite connected clusters of the system. In one dimension, the method is particularly simple due to the embedding structure of linear chain segments. As an example, we consider the ground-state energy $E_0$, which is an extensive and cluster-additive quantity. For two disconnected chain segments $A$ and $B$, additivity implies
\begin{equation}
E_0(A \cup B) = E_0(A) + E_0(B).
\end{equation}
This ensures that we only have to calculate contributions from connected subclusters. To avoid overcounting terms already present in smaller clusters, the reduced contribution $\widetilde{E}_0(C)$ of a connected cluster $C$ is defined recursively as
\begin{equation}
\widetilde{E}_0(C) = E_0(C) - \sum_{C' \subset C} \widetilde{E}_0(C'),
\label{eq:nlce_recursive}
\end{equation}
where the sum runs over all connected proper subclusters $C' \subset C$. For the chain geometry, the subclusters simplify to linear chain segments of increasing length. The recursion in Eq.~\eqref{eq:nlce_recursive} then leads to a telescopic structure, and the energy per site $e_0=E_0/N$ in the thermodynamic limit can be approximated by
\begin{equation}
e_0^{(N)} = E_0(C_{N+1}) - E_0(C_{N}),
\label{eq:nlcefinal}
\end{equation}
where $C_{N+1}$ and $C_{N}$ denote linear clusters of size $N+1$ and $N$, respectively. This difference yields the energy per site in the thermodynamic limit, with correlations included up to range $N$. By increasing the maximum cluster size $N$, one systematically incorporates longer-range correlations until convergence is achieved. A more detailed derivation of this formula can be found in \autoref{sec:nlcederivation}. For systems with finite correlation length, the series converges rapidly, allowing an accurate estimate of the ground-state energy density in the thermodynamic limit.

\subsection{DMRG}

To obtain the ground-state energy in the thermodynamic limit via the NLCE method described above, it is necessary to compute the ground-state energies of two finite clusters of sizes $N$ and $N+1$. In our NLCE+DMRG approach these cluster Hamiltonians are solved using the DMRG algorithm as implemented in the \texttt{ITensor} library~\cite{Fishman2020ITensor}. 

DMRG is a variational algorithm based on the matrix product state (MPS) ansatz, which provides an efficient representation of quantum many-body wave functions in one spatial dimension~\cite{White1992,White1993}. For a comprehensive review, see Ref.~\cite{Schollwock2011}. The algorithm proceeds by iteratively optimizing local tensors to minimize the ground-state energy, while truncating the Hilbert space according to the most significant eigenstates of the reduced density matrix. During this process, the bond dimension, controlling the amount of entanglement that can be represented, can be systematically increased to improve the accuracy of the result until convergence is reached. This adaptive control of the bond dimension makes DMRG both highly accurate and computationally efficient. 

The method is particularly powerful for one-dimensional systems, where the entanglement entropy of the ground state scales at most logarithmically with system size. 
This favorable scaling allows DMRG to achieve near machine-precision accuracy even for large systems and at quantum critical points. In our calculations, we employ the MPS and matrix product operator (MPO) framework provided by \texttt{ITensor}, which offers a robust and flexible implementation of DMRG. Convergence is controlled via the built-in \texttt{DMRGObserver} of \texttt{ITensor}~\cite{Fishman2020ITensor}. We use a strict truncation cutoff of $10^{-10}$ and an energy tolerance of $10^{-10}$. These parameters serve as default values throughout this work unless stated otherwise.

\subsection{Solving the self-consistency for ferromagnetic Ising couplings}

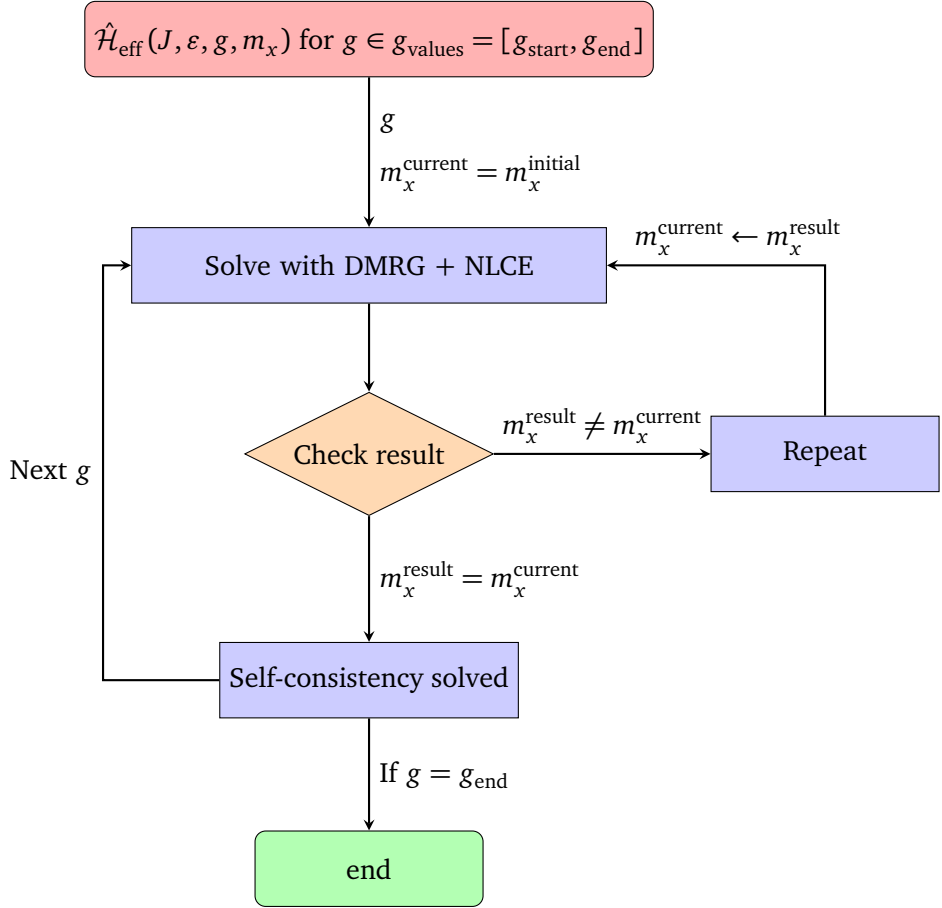
\begin{figure}[h]
    \centering
    \begin{tikzpicture}[node distance=2.5cm, scale=1, transform shape]
        \node (start) [start] {$\ham_{\text{eff}} \left(J,\varepsilon,g, m_x\right)$ 
        for $g \in g_{\text{values}}=\left[ g_{\text{start}},g_{\text{end}}\right]$ };
        \node (proc1) [process, text width = 6cm, align=center,below of=start, yshift=-0.5cm] {Solve with DMRG + NLCE};
        \node (dec1) [decision, below of=proc1] {Check result};
        \node (proc2a) [process, right of=dec1, xshift=3.5cm] {Repeat};
        \node (proc2b) [process, below of=dec1, yshift=-0.5cm] {Self-consistency solved};
        \node (stop) [stop, below of=proc2b, yshift=-0.0cm] {end};
        
        \draw [arrow] (start) -- node[right] {
          \shortstack[l]{
            $g$ \\[0.5em]
            $m_x^{\text{current}}=m_x^{\text{initial}}$
          }
        } (proc1);
        
        \draw [arrow] (proc1) -- (dec1);
        \draw [arrow] (dec1) -- node[above] {$m_x^{\text{result}} \neq m_x^{\text{current}}$} (proc2a);
        \draw [arrow] (dec1) -- node[right] {$m_x^{\text{result}} = m_x^{\text{current}}$} (proc2b);
        
        \draw [arrow] (proc2a.north) |- node[pos=0.7, above] {$m_x^{\text{current}}\leftarrow m_x^{\text{result}}$} (proc1.east);
        
        \draw[arrow] (proc2b) -- ++(-3.5,0) -- ++(0,5.5) node[pos=0.5, left] {Next $g$} -- (proc1.west);
        
        \draw [arrow] (proc2b) -- node[right] {If $g=g_{\text{end}}$} (stop);
    
    \end{tikzpicture}
    \caption{Visualization of the iterative procedure used to solve the self-consistency condition for ferromagnetic Ising interactions. The parameters $J$ and $\varepsilon$ are fixed, and the range of $g$ is specified. The procedure starts from an initial guess for $m_x$ and iterates until convergence. For the antiferromagnetic case, one has the same iterative procedure but an additional self-consistency loop for $m_s$.}
    \label{fig:selfconsistencyloop}
\end{figure}

Furthermore, a self-consistent determination of the transverse magnetization $m_x$ is required, which enters the effective matter Hamiltonian as a parameter. This self-consistency condition also has to be satisfied in the thermodynamic limit. To this end, we initialize the iterative procedure either in the strong-coupling regime (large $g$, where $m_x \approx 1/2$) or in the weak-coupling limit (small $g$, where $m_x \approx 0$), depending on the sweep direction. 
Performing both ``high'' and ``low'' sweeps enhance the convergence towards the global minimum of the total energy $E(m_x)$. For continuous transitions, both sweeps give identical results, while for first-order transitions they may differ close to the point of phase transition due to hysteresis behavior, allowing us to identify the ground state as the solution with the lower total energy.
Self-consistency is achieved by recursively inserting the output value of $m_x$ back into the Hamiltonian and solving the system until convergence is reached, i.e., when input and output agree. This iterative process is illustrated schematically in \autoref{fig:selfconsistencyloop}. To accelerate convergence, we employ a fixed-point iteration scheme based on Anderson acceleration~\cite{anderson1965iterative}.

Once the self-consistent solution has been obtained for a given value of $g$, the calculation is continued adiabatically by slightly increasing or decreasing $g$, using both the converged $m_x$ and the corresponding MPS wavefunctions as initial input. This adiabatic continuation allows us to efficiently map out the ground-state phase diagram over the entire range of coupling strengths.

\subsection{Solving the self-consistency for antiferromagnetic Ising couplings}
\label{sec:solveAf}

For antiferromagnetic Ising interactions, the magnetically ordered phase has alternatingly oriented nearest-neighbor spins. This implies that only even cluster sizes should be considered. Accordingly, the NLCE formula in Eq.~\eqref{eq:nlcefinal} must be adapted to a two-site unit cell. The corresponding expression reads
\begin{equation}
    e_0^{(N)} = \frac{E_0(C_{N+2}) - E_0(C_{N})}{2}.
\end{equation}

To characterize the antiferromagnetic order, we introduce the staggered magnetization $m_s$ along the $z$-direction as the order parameter: 
\begin{equation}
    m_s = \frac{1}{2N}\sum_{i=1}^N(-1)^i\langle \hat{\sigma}_i^z\rangle\,.
\end{equation}
Due to the underlying $\mathbb{Z}_2$ symmetry, one has two symmetry-equivalent ground-state sectors in the thermodynamic limit which couple on finite clusters. To set up an NLCE within a symmetry-broken phase, it is important to fix the ground-state sector on all clusters in order to obtain a consistent scheme.
This is implemented by adding the boundary field 
\begin{equation}
    \ham_{\text{env}} = 2\,J m_s \hat{\sigma}_1^z - 2\,J m_s \hat{\sigma}_N^z
    \label{eq:boundary_field}
\end{equation}
on each chain segment acting on both edge spins. Physically, this boundary field originates from the Ising coupling of the first and last spin to the ordered environment which is taken into account on the mean-field level.
The prefactor of 2 is chosen such that, in the fully antiferromagnetically ordered state with $m_s = 1/2$, the resulting boundary field corresponds to a coupling of strength $J$. While this choice is not essential, it provides a convenient normalization, which is consistent with the Ising interaction.

This environmental field breaks the $\mathbb{Z}_2$ symmetry and selects one of the two degenerate antiferromagnetically ordered states. On finite clusters, this induces a significant larger gap between the nearly degenerate eigenstates, proportional to the boundary field strength, which improves numerical convergence. This boundary field is determined self-consistently~\cite{ixert2016nonperturbative} giving rise to a second self-consistency condition $m_s$. While both $m_x$ and $m_s$ must be determined self-consistently, their roles in the Hamiltonian differ: $m_x$ enters as a global transverse field acting uniformly on all spins [Eq.~\eqref{eq:Heff}], whereas $m_s$ appears only in the boundary fields [Eq.~\eqref{eq:boundary_field}] that couple the finite cluster to its antiferromagnetically ordered environment. As a result, the effective Hamiltonian now depends on two coupled self-consistent parameters, $m_x$ and $m_s$, which must be determined simultaneously to obtain convergence.

\section{Results}
In the following we present the results obtained by the NLCE+DMRG approach for the ground-state properties of the ferromagnetic and antiferromagnetic Dicke-Ising chain in the thermodynamic limit.

\subsection{Ferromagnetic Dicke-Ising chain}
\label{sec:ferro}
In this subsection  we investigate the Dicke-Ising chain with ferromagnetic Ising interactions (\(J<0\)). We quantitatively determine the quantum phase diagram and locate the multicritical point where the quantum phase transition between the non-superradiant and superradiant phase changes from second to first order.

To do so, a key question is how to determine the order of the quantum phase transition in the ferromagnetic Dicke-Ising chain and identify where it changes as a function of the system parameters. This can be done by analyzing the transition point in the spin-photon coupling \(g/\omega_c\) and comparing it to the 
value of the Dicke-type polariton condensation given exactly by mean-field theory \cite{Zhang2014, Schellenberger_2024}. 
Indeed, in the weak-coupling limit,
$
g \ll \min\{\vert J\vert, \omega_c, \varepsilon\},
$
the photon field has no effect on the ferromagnetic ground state in the thermodynamic limit. In this regime, the system remains in its unperturbed spin configuration, namely the two degenerate ground states with all spins polarized either along \( +z \) or along \( -z \), with a vanishing photon occupation.
In the ferromagnetic phase, the ground-state energy per site is
\begin{equation}
    \frac{E_0}{N} = -\frac{Jc}{2} - \varepsilon,
\end{equation}
where \(c=2\) denotes the lattice connectivity of the chain, i.e., the number of nearest neighbors per site.

Mean-field theory \cite{Zhang2014, Schellenberger_2024} predicts a continuous phase transition at the critical spin-photon coupling
\begin{equation}
    g_{\text{crit}}^{\text{MF}} = \sqrt{2\varepsilon + 2cJ}.
\end{equation}

To determine the order of the transition, we compute the ground-state energy as a function of \(g\) and compare its value at the mean-field critical point \(g = g_{\text{crit}}^{\text{MF}}\) to the exact ground-state energy per site in the weak-coupling magnetically ordered phase as already done in Ref.~\cite{Koziol2025}. Two cases arise:

\begin{itemize}
    \item If the numerically obtained ground-state energy at the transition is lower than the mean-field value, the transition occurs already at a smaller coupling \(g < g_{\text{crit}}^{\text{MF}}\). In this case the transition is \textit{first-order}.

    \item If the ground-state energy matches the mean-field prediction at \(g = g_{\text{crit}}^{\text{MF}}\), the transition is a \textit{second-order} transition.
\end{itemize}

This energy-based criterion provides a simple and effective method for identifying the order of the quantum phase transition and assessing the validity of the mean-field description across parameter regimes. In addition, we compute the corresponding order parameters within our NLCE+DMRG approach to verify self-consistency and to confirm the expected behavior: discontinuities in the transverse magnetization accompany first-order transitions, while a continuous onset is observed at second-order ones.
\subsubsection{Determination of the multicritical point}

\begin{figure}[t]
    \centering
    \includegraphics{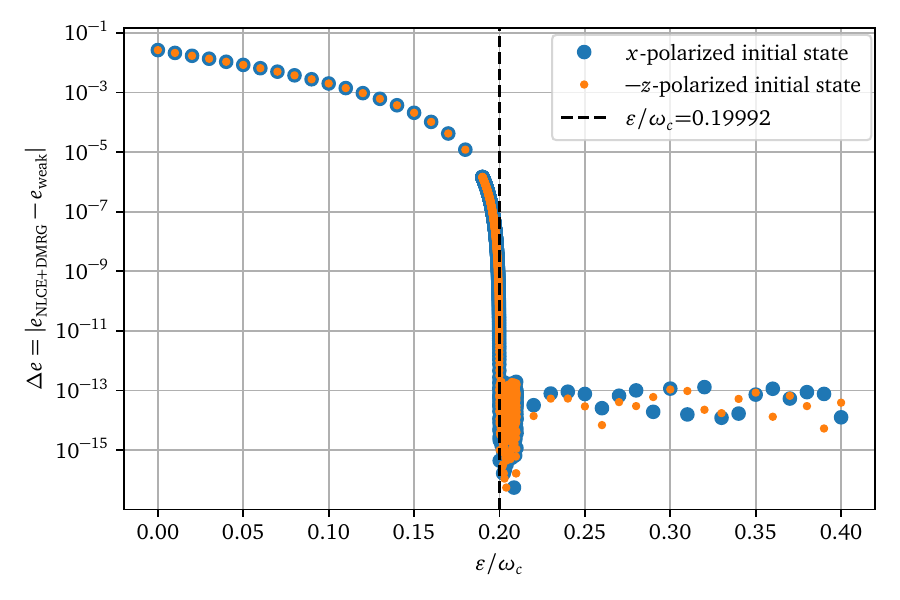}

    \caption{
    Absolute value of the energy difference \(\Delta e = \left| e_{\mathrm{NLCE+DMRG}} - e_{\mathrm{weak}}\right|\) obtained from NLCE+DMRG at the mean-field critical coupling \(g_{\mathrm{crit}}\).  Calculations are performed on clusters of size \(N=100\) and \(N=101\). Finite values \(\Delta e >0\) signal a first-order transition, while \(\Delta e \approx 0\) corresponds to a continuous transition.  
    The curves show results starting from \(x\)-polarized and \(-z\)-polarized initial states, plotted in the same panel to highlight their near-perfect agreement over a wide parameter range.  
    Since the data are shown on a logarithmic scale, values below the numerical accuracy (\(\sim 10^{-12}\)) are effectively rendered as zero. The vertical line marks the first data point at which the two curves become numerically distinguishable, indicating that the numerical accuracy is reached and providing a practical criterion for identifying the multicritical point.
    DMRG convergence thresholds are \(10^{-14}\) in energy and \(10^{-13}\) for the self-consistency loop.
    }
    \label{fig:deltaE_ferro}
\end{figure}

We now apply this criterion to the ferromagnetic Dicke-Ising chain for \(J=-0.2\).
To detect the change in transition order, we compute
\[
\Delta e = e_{\mathrm{NLCE+DMRG}} - e_{\mathrm{weak}},
\]
where \(e_{\mathrm{NLCE+DMRG}}\) is the numerically determined ground-state energy per site at the mean-field critical point \(g=g_{\text{crit}}^{\text{MF}}\), and \(e_{\mathrm{weak}}\) is the exact ground-state energy per site in the weak-coupling magnetically ordered phase. The results are shown in \autoref{fig:deltaE_ferro}.

We perform the DMRG calculation twice: once using a \(-z\)-polarized MPS and once using a \(+x\)-polarized MPS as initial states. These product states correspond to the exact ground states in the weak- and strong-coupling limits, respectively, and are therefore natural MPS input states for our DMRG procedure. Both ansätze yield identical results, further confirming the robustness of our findings.

For small longitudinal fields \( \varepsilon/\omega_c \lesssim 0.2 \), we find \( \Delta e > 0 \), showing that the phase transition occurs at a coupling smaller than the mean-field prediction, thus marking it as a first-order transition. 
At $\varepsilon/\omega_c = 0.19992 \pm 0.00005$, the energy difference has already decreased to values of order $10^{-12}$, while the curves obtained from the $-x$- and $+z$-polarized initial states still coincide to even higher precision. 
At this coupling, $\Delta e$ reaches the resolution limit of our calculation and becomes numerically indistinguishable from zero, meaning that the ground-state energy matches the mean-field prediction within our achievable accuracy. 
We therefore identify this point as the multicritical point. For larger longitudinal fields, $\Delta e$ remains numerically zero, consistent with a second-order transition beyond the multicritical point.
Based on this analysis, we identify \(\varepsilon/\omega_c = 0.19992 \pm 0.00005\) as the field strength of the multicritical point at which the quantum phase transition between non-superradiant and superradiant phases changes from first to second order.

\subsubsection{Complete phase diagram}
\begin{figure}[t]
    \centering
    \includegraphics{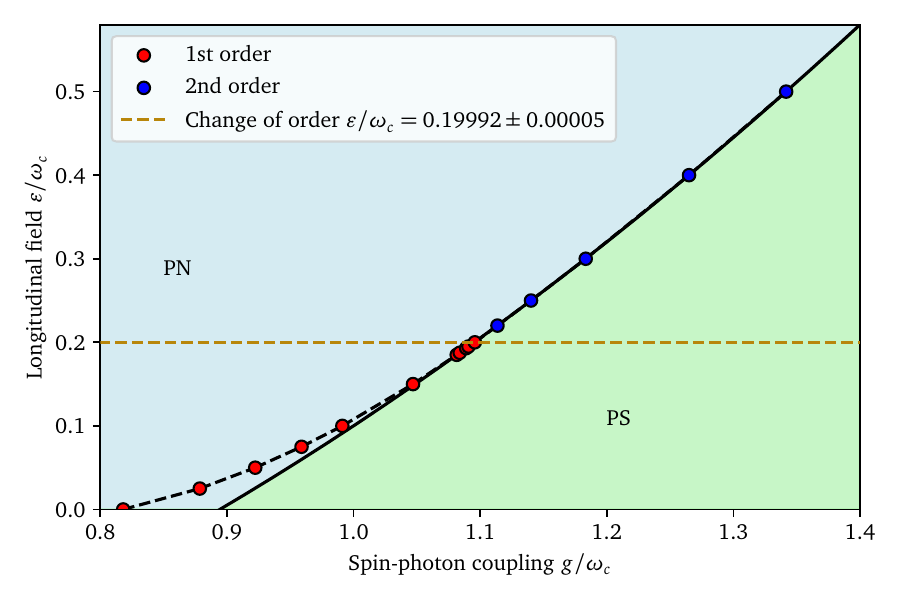}
    \caption{
    Phase diagram of the Dicke-Ising chain with ferromagnetic interactions (\(J=-0.2\)).  
    Red dots correspond to first-order transitions; blue dots indicate continuous transitions.  
    The black curve is the mean-field PN-PS boundary.  
    The brown dashed line marks the multicritical point at \(\varepsilon/\omega_c = 0.19992 \pm 0.00005\).}
    \label{fig:ferro_phase}
\end{figure}

After locating the multicritical point, we map out the full ferromagnetic phase as shown in \autoref{fig:ferro_phase}.  
Red points mark first-order transitions extracted from NLCE+DMRG, while blue points indicate continuous transitions.  
The mean-field phase boundary separating the paramagnetic normal (PN) and paramagnetic superradiant (PS) phases is shown as a black line, and the multicritical field \(\varepsilon/\omega_c = 0.19992 \pm 0.00005\) is highlighted by a brown dashed line.

At small longitudinal fields, the first-order boundary lies significantly below the mean-field prediction, while at larger fields the boundaries coincide within numerical precision with the mean-field ciritical value.  
These results indicate that the ferromagnetic Dicke-Ising chain exhibits a change in transition order controlled by the ratio $\varepsilon/|J|$ that we have explicitly demonstrated for $J=-0.2$.
Further, the NLCE+DMRG approach reliably captures both first-order and second-order regimes.

In summary, our NLCE+DMRG approach provides a high-precision characterization of the ferromagnetic transition directly in the thermodynamic limit. 
QMC simulations~\cite{langheld} yield highly accurate results on finite systems but relies on finite-size scaling to access the thermodynamic limit; its estimate near \( J \approx \varepsilon \) for the multicritical point agrees well with our findings. 
In contrast, recent finite-size DMRG studies~\cite{mendoncca2025, comment2025} do not resolve the multicritical point, which emerges clearly in our thermodynamic-limit analysis.

\subsection{Antiferromagnetic Dicke-Ising chain}
\label{sec:af}

Before discussing the antiferromagnetic phase diagram at finite $g$, it is instructive to consider the limit $g = 0$, where the effective matter model reduces to the classical antiferromagnetic Ising chain in a longitudinal field,
\begin{equation}
    \ham = \varepsilon \sum_i \sigma_i^z + J \sum_i \sigma_i^z \sigma_{i+1}^z \,.
\end{equation}
A direct energy comparison between the antiferromagnetically ordered state ($E/N = -J$) and the polarized state ($E/N = J - \varepsilon$) yields a first-order transition at $\varepsilon/J = 2$, marked by a kink in the ground-state energy and a discontinuous jump in the magnetization. The classical nature of the model leads to an extensive ground-state degeneracy. For finite $g$, quantum fluctuations lift this degeneracy and the phase boundaries separate.

For antiferromagnetic Ising interactions (\(J > 0\))  the Dicke-Ising model exhibits qualitatively different behavior compared to the ferromagnetic case. As shown in the mean-field phase diagram in \autoref{fig:mf_phase_gJ}, an additional antiferromagnetic superradiant (AS) phase with simultaneous order in light and matter emerges at intermediate coupling strengths. It combines antiferromagnetic order with a finite cavity-field expectation value, thus breaking both translational invariance and parity symmetries.
QMC simulations~\cite{langheld} in one and two dimensions confirmed the presence of this intermediate AS. Recent finite-size DMRG studies~\cite{mendoncca2025}, however, reported its absence in one dimension, ignoring and contradicting these works, as discussed in an associated comment~\cite{comment2025}. Our NLCE+DMRG approach, which systematically eliminates finite-size effects, quantitatively confirms the existence of the AS phase in the Dicke-Ising chain without any doubt and reveals its narrow extension (\(\Delta g/\omega_c \approx 0.02\)). Further, our approach allows a microscopic characterization of the AS phase, which emerges from the antiferromagnetic normal phase through the softening of the Dicke-type polariton mode induced by the self-consistent cavity field.

\subsubsection{Numerical determination of the phase boundaries with NLCE+DMRG}
\label{subsec:af_numerical}

We employ two complementary numerical protocols within the NLCE+DMRG framework as introduced in \autoref{sec:solveAf}:  
(i) a sweep in the spin-photon coupling \(g\) at fixed longitudinal field \(\varepsilon\), and  
(ii) a sweep in \(\varepsilon\) at fixed \(g\).  
Both sweeps are performed adiabatically in two directions, starting from large and from small parameter values, to account for possible hysteresis, which is characteristic for first-order transitions.  
The \(g\)-sweep directly resolves the AN-AS-PS structure, while the \(\varepsilon\)-sweep additionally accesses the PN phase and enables a direct comparison with QMC data \cite{langheld}.

Throughout this subsubsection, we first analyze the representative point $J = 0.2$, $\varepsilon = 0.3$ using the $g$-sweep, which is visualized in \autoref{fig:mf_phase_gJ}, before turning to the $\varepsilon$-sweep at fixed $g = 0.52$.

\subsubsection*{Sweep in \texorpdfstring{\(g\)}{g}}
\begin{figure}[t]
    \centering
    \includegraphics{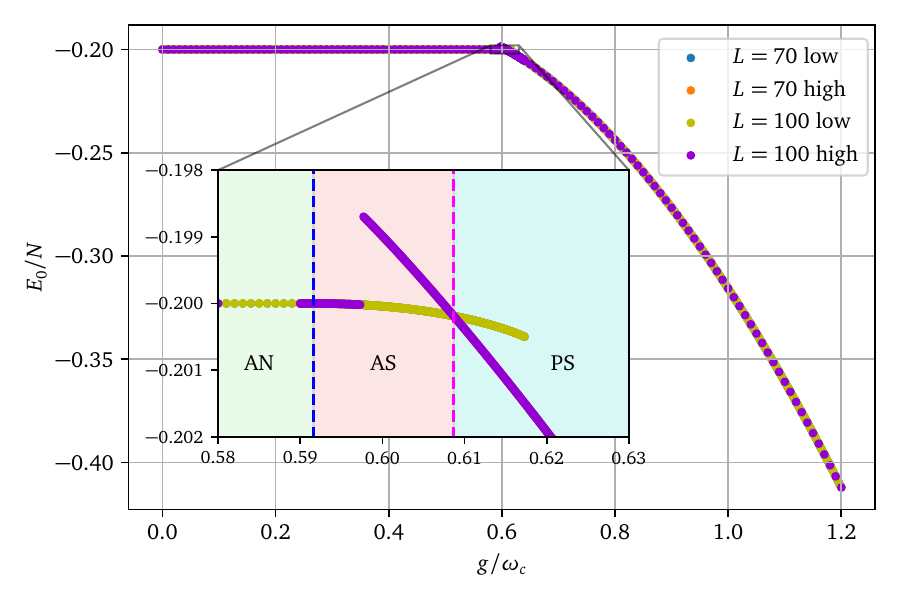}
    \caption{Ground-state energy per site $E_0/N$ as a function of \(g/\omega_c\) for the antiferromagnetic Dicke-Ising chain with \(J = 0.2\), \(\varepsilon = 0.3\).  
    The AN-AS transition is continuous, whereas the AS-PS transition shows a clear first-order signature through a discontinuity between increasing and decreasing sweeps.  
    The inset magnifies the transition region.}
    \label{fig:af_energy}
\end{figure}

\autoref{fig:af_energy} shows the ground-state energy per site $E_0/N$ as a function of \(g/\omega_c\) for increasing and decreasing sweeps.  
The AN-AS transition occurs smoothly without an energy discontinuity, consistent with a continuous transition. By contrast, the AS-PS transition displays a clear first-order character: the two sweeps yield different branches in the transition region, and the energy curves cross.  
The dashed vertical lines indicate the positions of both transitions. 

\begin{figure}[t]
    \centering
    \includegraphics{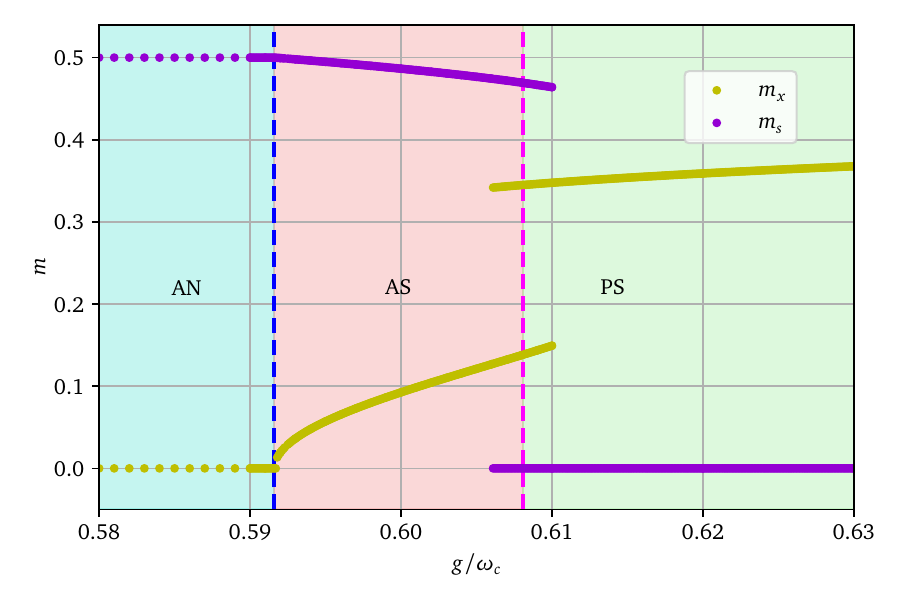}
    \caption{Staggered magnetization \(m_s\) and transverse magnetization \(m_x\) as functions of \(g/\omega_c\) for \(J = 0.2\), \(\varepsilon = 0.3\).  
    In the AS phase both order parameters are finite, signaling simultaneous breaking of spin-flip and parity symmetries.}
    \label{fig:af_mags}
\end{figure}

To further characterize the AS phase, we evaluate two order parameters:  
(i) the staggered magnetization \(m_s\), signaling antiferromagnetic order, and  
(ii) the transverse magnetization \(m_x\), which measures the breaking of parity symmetry through the coupling to the photon field.  
As shown in \autoref{fig:af_mags}, both order parameters are finite in the AS phase, demonstrating simultaneous breaking of translational invariance and parity symmetries.  
In the AN phase only \(m_s\) remains finite, while in the PS phase only \(m_x\) survives.

These results confirm that the AS phase is a genuine intermediate phase, bounded by a continuous AN-AS transition and a first-order AS-PS transition, an effect specific to antiferromagnetic interactions and absent in the ferromagnetic Dicke-Ising chain.

\subsubsection*{Sweep in \texorpdfstring{\(\varepsilon\)}{epsilon}}

To compare directly with the QMC results of Langheld \emph{et al.}~\cite{langheld}, we perform an independent sweep in the longitudinal field \(\varepsilon\) while keeping the spin-photon coupling fixed at \(g = 0.52\), following the protocol in \autoref{sec:solveAf}. 
This sweep traverses all four phases AN, AS, PS, and PN of the antiferromagnetic Dicke-Ising chain.

\begin{figure}[t]
    \centering
    \includegraphics{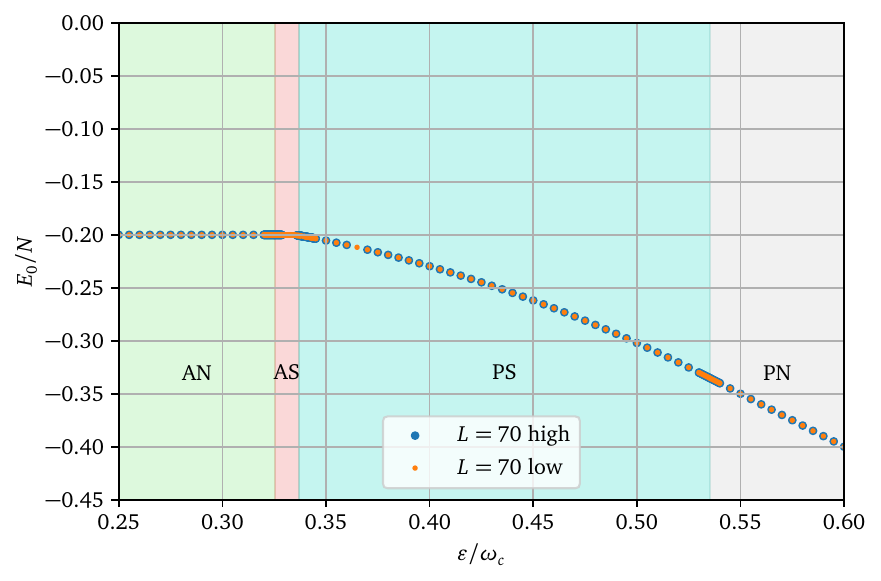}
    \caption{Ground-state energy per site $E_0/N$ as a function of \(\varepsilon/\omega_c\) for \(J = 0.2\), \(g = 0.52\).  
    The AN-AS transition is continuous, while the AS-PS transition shows a first-order signature.  
    The PS-PN transition is continuous.}
    \label{fig:af_energyconstg}
\end{figure}

\begin{figure}[t]
    \centering
    \includegraphics{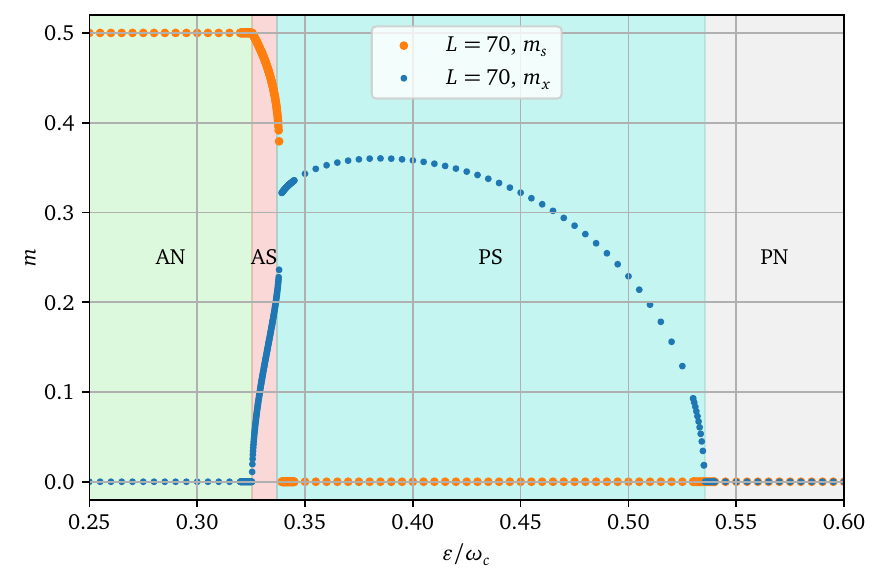}
    \caption{Staggered magnetization \(m_s\) and transverse magnetization \(m_x\) as functions of \(\varepsilon/\omega_c\) for \(J = 0.2\), \(g = 0.52\).}
    \label{fig:af_magsconstg}
\end{figure}

\autoref{fig:af_energyconstg} shows the ground-state energy per site $E_0/N$ as a function of \(\varepsilon/\omega_c\).  
The AN-AS boundary remains continuous, while the AS-PS transition is clearly first order. The PS-PN transition at larger $\varepsilon$ is continuous, as expected.
The corresponding order parameters in \autoref{fig:af_magsconstg} confirm this picture:  
\(m_s\) and \(m_x\) behave analogously to the \(g\)-sweep and both vanish in the PN phase, where no magnetic or transverse order is present.

The extracted phase boundaries are summarized in \autoref{tab:phase_boundaries}.  
We find excellent agreement with the QMC data~\cite{langhelddata}, confirming both the existence and the extent of the AS phase. Furthermore, the critical values agree with the exact mean-field predictions at the AN-AS and PS-PN transitions.

\begin{table}[t]
    \centering
    \begin{tabular}{lccc}
        \hline\hline
        \textbf{Phase boundary} & \textbf{NLCE+DMRG} & \textbf{QMC} & \textbf{Mean-field (exact)} \\
        \hline
        AN - AS & 0.3254 & 0.3254& 0.3254\\
        AS - PS & 0.3369 & 0.3365& --\\
        PS - PN & 0.5355 & --& 0.5352\\
        \hline\hline
    \end{tabular}
    \caption{Comparison of phase boundaries obtained from NLCE+DMRG, QMC~\cite{langhelddata}  and exact mean-field values~\cite{Zhang2014,Schellenberger_2024} for \(J = 0.2\), \(g = 0.52\) in units of \(\varepsilon/\omega_c\). The NLCE+DMRG boundaries have a numerical uncertainty of \(\pm 0.0005\).}
    \label{tab:phase_boundaries}
\end{table}

Overall, our approach works robustly for antiferromagnetic Ising interactions.  
The two-site unit cell and the environmental field (see Eq.~\eqref{eq:boundary_field}) used in the NLCE allow us to resolve the staggered magnetization and reliably identify all phase boundaries.  
We clearly observe the intermediate AS phase, in quantitative agreement with QMC~\cite{langheld} and in contrast to recent finite-size DMRG studies~\cite{mendoncca2025}.

\subsubsection{Many-body ground state of the AS phase}
\label{subsec:af_microscopic}
The effective matter Hamiltonian framework allows us to go beyond order-parameter characterization and to identify what the AS phase microscopically is: the many-body ground state of the effective matter Hamiltonian \eqref{eq:Heff} at the self-consistent value \(m_x > 0\).

The emergence of the AS phase can be understood through the energy landscape \(E(m_x)\) in the extended parameter space. 
In the AN phase, the global minimum lies at \(m_x = 0\), corresponding to an antiferromagnetically ordered state without photon condensate. 
As the spin-photon coupling \(g\) increases, Dicke-type polariton condensation occurs. The associated condensation mode couples uniformly to all spins (zero momentum) and as a consequence cannot destroy the antiferromagnetic order. 
The polariton condensation out of the AN phase therefore can only lead to the AS phase and not to the PS phase. 
A direct continuous AN-PS transition is also excluded since it would require a quantum phase transition between two symmetry-broken phases. 
After the transition, the system resides in a state with finite \(m_x > 0\) while retaining antiferromagnetic order. 
The AS ground state is adiabatically connected, in the extended parameter space \((g, \varepsilon, m_x)\), to the antiferromagnetically ordered ground state of the transverse-field Ising model with longitudinal field. 
The photon occupation in the original Dicke-Ising model is directly given by the transverse magnetization \(m_x\).

The energy landscape perspective also clarifies why intermediate superradiant phases with simultaneous magnetic order do not arise generically in the effective Hamiltonian \eqref{eq:Heff}. 
While the latter admits such states, they do not have to be the global minimum at given parameter values. Instead, the ground states of non-superradiant magnetically ordered phases and the PS phase can have lower energies, leading to a first-order transition between the two. The emergence of an intermediate superradiant magnetically ordered phase can be understood as a reshaping of the energy landscape by the longitudinal field \(\varepsilon\) such that the magnetically ordered superradiant ground state becomes energetically favorable.

Our numerical results confirm these considerations.  
In particular, the AN-AS transition is continuous and coincides with the mean-field phase boundary \autoref{tab:phase_boundaries}. 
Following the analysis in Ref.~\cite{Schellenberger_2024}, agreement with the mean-field boundary identifies the mechanism as Dicke-type polariton condensation. 
The subsequent AS-PS transition is first order. 
Both findings are consistent with QMC~\cite{langheld}.

\section{Conclusion and outlook}
\label{sec:conclusion}

In this work, we have established a thermodynamic-limit description of the Dicke-Ising model by mapping it to a self-consistent effective spin Hamiltonian, in which the cavity only enters via a self-consistent transverse field. This mapping is exact in the thermodynamic limit and allows us to access ground-state properties of the full light-matter system by solving a pure spin model. Building on this formulation, we combined the numerical linked-cluster expansion with DMRG (NLCE+DMRG) to obtain high-precision results for the Dicke-Ising chain in the thermodynamic limit.

For ferromagnetic Ising interactions, we demonstrated that the Dicke-Ising chain exhibits a change of the order of the superradiant quantum phase transition. Using an energy-based criterion at the mean-field critical coupling, we identified \(\varepsilon/\omega_c = 0.19992 \pm 0.00005\) as the
multicritical point for \(J=-0.2\). For smaller fields, the superradiant transition is first order and occurs at couplings smaller than predicted by mean-field theory \cite{Zhang2014}. For larger $\epsilon/\omega_c$ the transition is continuous and the NLCE+DMRG results coincide, as expected, with the mean-field boundary within numerical accuracy. 

For antiferromagnetic Ising interactions, we clarify the microscopic origin and the phase boundaries of the antiferromagnetic superradiant (AS) phase in one dimension. Within the effective matter Hamiltonian picture, the AS phase emerges from the antiferromagnetic normal state via a self-consistent cavity-induced transverse field. Our NLCE+DMRG calculations demonstrate that the AS phase features simultaneous order in the spin and photon sectors, and is bounded by a continuous AN-AS transition and a first-order AS-PS transition. We find that this phase occupies only a very narrow window in the coupling \(g\), in quantitative agreement with QMC results \cite{langheld}, thereby explaining why the finite-size DMRG approach of Ref.~\cite{mendoncca2025} failed to resolve it, as discussed in Ref.~\cite{comment2025}.

Methodologically, our work highlights the effectiveness of combining the exact effective-Hamiltonian mapping with NLCE+DMRG in one dimension. The approach is nonperturbative, directly formulated in the thermodynamic limit, and exploits the particular strength of DMRG in one spatial dimension, where entanglement is efficiently captured by matrix product states. As a result, we can reach large effective system sizes with controlled errors, making the method particularly powerful for one-dimensional cavity-coupled spin models.

Let us mention that our approach and the QMC algorithm in Ref.~\cite{langheld} are complementary with respect to finite-size effects. NLCE+DMRG applied on the effective matter Hamiltonian operates directly in the thermodynamic limit, achieving high precision results without further finite-size extrapolation for the Dicke-Ising chain. Let us stress that for other matter Hamiltonians with quantum-critical points or gapless phases also NLCE usually needs an extrapolation in the size of the used cluster to extract the physical properties of the system. However, finite-size effects are always distinct and reduced compared to treatments of the light-matter Hamiltonian on finite systems like the QMC algorithm in Ref.~\cite{langheld} or the DMRG calculations in Ref.~\cite{Sur2025}. In particular, the QMC algorithm solves large finite systems up to several thousands of spins of the Dicke-Ising model with numerical precision, making it the method of choice for studying finite-size systems in cavity QED platforms. A systematic analytical treatment of $1/N$ corrections, building on the structure of the remainder term in the displaced Hamiltonian, could bridge these approaches and clarify how the phase structure emerges with increasing system size.

Several directions for future work are apparent. On the methodological side, extending the NLCE+DMRG scheme to higher-dimensional and frustrated lattices as well as to other matter-matter interactions, would allow one to explore how dimensionality and geometric frustration affect phases with simultaneous order \cite{Koziol2025, Schuler2020}, such as the AS phase, and what kind of other quantum phases can be stabilized. In higher dimensions, tensor-network methods based on projected entangled pair states (PEPS)~\cite{verstraete2004} provide a natural replacement for DMRG and could be combined with the effective-Hamiltonian mapping. 
On the conceptual side, analyzing finite-temperature properties and dynamical responses within the effective matter Hamiltonian framework would provide a more direct connection to experiments. This also includes the study of thermal phase transitions \cite{gammelmark2011, otake2025}.

\section*{Acknowledgements}

\paragraph{Funding information}
The authors gratefully acknowledge the support by the Deutsche For\-schungsgemeinschaft (DFG, German Research Foundation) -- Project-ID 429529648 -- TRR 306 \mbox{QuCoLiMa} ("Quantum Cooperativity of Light and Matter") and the Munich Quantum Valley, which is supported by the Bavarian state government with funds from the Hightech Agenda Bayern Plus.

\addtocontents{toc}{\protect\setcounter{tocdepth}{1}}

\begin{appendix}

\numberwithin{equation}{section}

\section{Derivation of the 1D NLCE Formula}
\label{sec:nlcederivation}
To illustrate how the numerical linked-cluster expansion (NLCE) simplifies in one-dimensional systems, we derive a compact expression for computing extensive observables in the thermodynamic limit using only the largest clusters, as presented in the main text in Eq.~\eqref{eq:nlcefinal}.
Let \(\mathcal{M}(C_i)\) be the value of a cluster-additive observable \(\mathcal{M}\), such as the ground-state energy, evaluated on a chain segment \(C_i\) of size \(i\). The reduced contribution \(\widetilde{\mathcal{M}}(C_N)\) of a cluster \(C_N\) is defined by subtracting overlapping contributions from smaller clusters:
\begin{equation}
\widetilde{\mathcal{M}}(C_N) = \mathcal{M}(C_N) - \sum_{i=1}^{N-1} a_i\, \widetilde{\mathcal{M}}(C_i),
\label{eq:nlce_reduced}
\end{equation}
where \(a_i\) is the embedding factor counting how often a cluster of size \(i\) fits into one of size \(N\). In open chains, this is simply \(a_i = N - i + 1\).
Now, consider the cluster of size \(N+1\). Its reduced contribution is
\begin{equation}
\widetilde{\mathcal{M}}(C_{N+1}) = \mathcal{M}(C_{N+1}) - \sum_{i=1}^{N} b_i\, \widetilde{\mathcal{M}}(C_i),
\end{equation}
with \(b_i = N - i + 2\), since one more embedding is possible than in the \(N\)-site case. 
The reduced contribution $\widetilde{\mathcal{M}}(C_{N+1})$ can then be expressed as
\begin{equation}
\widetilde{\mathcal{M}}(C_{N+1}) = \mathcal{M}(C_{N+1}) - \sum_{i=1}^{N-1} b_i\, \widetilde{\mathcal{M}}(C_i) - 2\,\widetilde{\mathcal{M}}(C_N).
\end{equation}
Subtracting \(\widetilde{\mathcal{M}}(C_N)\) from the above yields 
\begin{equation}
\widetilde{\mathcal{M}}(C_{N+1}) - \widetilde{\mathcal{M}}(C_N) = \mathcal{M}(C_{N+1}) - \mathcal{M}(C_N) - \sum_{i=1}^{N-1} \widetilde{\mathcal{M}}(C_i) - 2\,\widetilde{\mathcal{M}}(C_N).
\label{eq:nlce_difference}
\end{equation}
To reconstruct the observable in the thermodynamic limit up to cluster of size \(N+1\), we sum over all reduced contributions:
\begin{align}
\mathcal{M}(\mathcal{L}_\infty) &= \sum_{i=1}^{N+1} \widetilde{\mathcal{M}}(C_i) \nonumber \\
&= \sum_{i=1}^{N} \widetilde{\mathcal{M}}(C_i) + \widetilde{\mathcal{M}}(C_{N+1}) \nonumber \\
&= \sum_{i=1}^{N} \widetilde{\mathcal{M}}(C_i) + \mathcal{M}(C_{N+1}) - \mathcal{M}(C_N) - \sum_{i=1}^{N} \widetilde{\mathcal{M}}(C_i) \nonumber \\
&= \mathcal{M}(C_{N+1}) - \mathcal{M}(C_N),
\label{eq:nlce_final}
\end{align}
which is the desired NLCE formula for 1D systems.

\end{appendix}





\bibliography{SciPost_Example_BiBTeX_File.bib}

@article{Shapiro2025,
  title = {Digital-analog simulations of Schr\"odinger cat states in the Dicke-Ising model},
  author = {Shapiro, Dmitriy S. and Weber, Yannik and Bode, Tim and Wilhelm, Frank K. and Bagrets, Dmitry},
  journal = {Phys. Rev. A},
  volume = {112},
  issue = {4},
  pages = {042412},
  numpages = {15},
  year = {2025},
  month = {Oct},
  publisher = {American Physical Society}, 
  doi = {10.1103/wbp6-y3vd},
  url = {https://link.aps.org/doi/10.1103/wbp6-y3vd}
}

@Article{Schuler2020,
	title={{The vacua of dipolar cavity quantum electrodynamics}},
	author={Michael Schuler and Daniele De Bernardis and Andreas M. Läuchli and Peter Rabl},
	journal={SciPost Phys.},
	volume={9},
	pages={066},
	year={2020},
	publisher={SciPost},
	doi={10.21468/SciPostPhys.9.5.066},
	url={https://scipost.org/10.21468/SciPostPhys.9.5.066},
}

@misc{otake2025,
      title={{Exactly Solvable Phase Transition in a Cavity-Coupled 1D Ising Chain}}, 
      author={Shuntaro Otake and Motoaki Bamba},
      year={2025},
      eprint={2507.01486},
      archivePrefix={arXiv},
      primaryClass={cond-mat.stat-mech},
      url={https://arxiv.org/abs/2507.01486}, 
}

@article{Koziol2025,
  title = {Melting of devil's staircases in the long-range Dicke-Ising model},
  author = {Koziol, Jan Alexander and Langheld, Anja and Schmidt, Kai Phillip},
  journal = {Phys. Rev. B},
  volume = {111},
  issue = {22},
  pages = {224427},
  numpages = {19},
  year = {2025},
  month = {Jun},
  publisher = {American Physical Society},
  doi = {10.1103/syps-9r7r},
  url = {https://link.aps.org/doi/10.1103/syps-9r7r}
}

@article{langheld,
  title = {Quantum phase diagrams of Dicke-Ising models by a wormhole algorithm},
  author = {Langheld, Anja and H\"ormann, Max and Schmidt, Kai Phillip},
  journal = {Phys. Rev. B},
  volume = {112},
  issue = {16},
  pages = {L161123},
  numpages = {7},
  year = {2025},
  month = {Oct},
  publisher = {American Physical Society},
  doi = {10.1103/lcvj-ksct},
  url = {https://link.aps.org/doi/10.1103/lcvj-ksct}
}

@article{rohn,
  title = {Ising model in a light-induced quantized transverse field},
  author = {Rohn, Jonas and H\"ormann, Max and Genes, Claudiu and Schmidt, Kai Phillip},
  journal = {Phys. Rev. Res.},
  volume = {2},
  issue = {2},
  pages = {023131},
  numpages = {15},
  year = {2020},
  month = {May},
  publisher = {American Physical Society},
  doi = {10.1103/PhysRevResearch.2.023131},
  url = {https://link.aps.org/doi/10.1103/PhysRevResearch.2.023131}
}

@article{Zhang2014,
  author    = {Yuanwei Zhang and Lixian Yu and J. -Q. Liang and Gang Chen and Suotang Jia and Franco Nori},
  title     = {Quantum phases in circuit QED with a superconducting qubit array},
  journal   = {Scientific Reports},
  volume    = {4},
  number    = {1},
  pages     = {4083},
  year      = {2014},
  doi       = {10.1038/srep04083},
  url       = {https://doi.org/10.1038/srep04083},
  issn      = {2045-2322}
}

@article{Schellenberger_2024,
   title={{(Almost) everything is a Dicke model - Mapping non-superradiant correlated light-matter systems to the exactly solvable Dicke model}},
   volume={7},
   ISSN={2666-9366},
   url={http://dx.doi.org/10.21468/SciPostPhysCore.7.3.038},
   DOI={10.21468/scipostphyscore.7.3.038},
   number={3},
   journal={SciPost Physics Core},
   publisher={Stichting SciPost},
   author={Schellenberger, Andreas and Schmidt, Kai Phillip},
   year={2024},
   month=jul }

@article{dicke1954coherence,
  title = {Coherence in Spontaneous Radiation Processes},
  author = {Dicke, R. H.},
  journal = {Phys. Rev.},
  volume = {93},
  issue = {1},
  pages = {99--110},
  numpages = {0},
  year = {1954},
  month = {Jan},
  publisher = {American Physical Society},
  doi = {10.1103/PhysRev.93.99},
  url = {https://link.aps.org/doi/10.1103/PhysRev.93.99}
}

@misc{Sur2025,
      title={Amplified response of cavity-coupled quantum-critical systems}, 
      author={Shouvik Sur and Yiming Wang and Mounica Mahankali and Silke Paschen and Qimiao Si},
      year={2025},
      eprint={2509.26620},
      archivePrefix={arXiv},
      primaryClass={cond-mat.str-el},
      url={https://arxiv.org/abs/2509.26620}, 
}

@article{minimalcoupling1,
  title = {Phase Transitions, Two-Level Atoms, and the ${A}^{2}$ Term},
  author = {Rza\ifmmode \dot{z}\else \.{z}\fi{}ewski, K. and W\'odkiewicz, K. and \ifmmode \dot{Z}\else \.{Z}\fi{}akowicz, W.},
  journal = {Phys. Rev. Lett.},
  volume = {35},
  issue = {7},
  pages = {432--434},
  numpages = {0},
  year = {1975},
  month = {Aug},
  publisher = {American Physical Society},
  doi = {10.1103/PhysRevLett.35.432},
  url = {https://link.aps.org/doi/10.1103/PhysRevLett.35.432}
}

@article{minimalcoupling2,
  author    = {Anton Frisk Kockum and Adam Miranowicz and Simone De Liberato and Salvatore Savasta and Franco Nori},
  title     = {Ultrastrong coupling between light and matter},
  journal   = {Nature Reviews Physics},
  year      = {2019},
  volume    = {1},
  number    = {1},
  pages     = {19--40},
  doi       = {10.1038/s42254-018-0006-2},
  url       = {https://doi.org/10.1038/s42254-018-0006-2},
  issn      = {2522-5820}
}

@article{minimalcoupling3,
  title = {Gauge invariance of the Dicke and Hopfield models},
  author = {Garziano, Luigi and Settineri, Alessio and Di Stefano, Omar and Savasta, Salvatore and Nori, Franco},
  journal = {Phys. Rev. A},
  volume = {102},
  issue = {2},
  pages = {023718},
  numpages = {11},
  year = {2020},
  month = {Aug},
  publisher = {American Physical Society},
  doi = {10.1103/PhysRevA.102.023718},
  url = {https://link.aps.org/doi/10.1103/PhysRevA.102.023718}
}

@article{superrandiantMarquardt,
  title = {Superradiant Phase Transitions and the Standard Description of Circuit QED},
  author = {Viehmann, Oliver and von Delft, Jan and Marquardt, Florian},
  journal = {Phys. Rev. Lett.},
  volume = {107},
  issue = {11},
  pages = {113602},
  numpages = {5},
  year = {2011},
  month = {Sep},
  publisher = {American Physical Society},
  doi = {10.1103/PhysRevLett.107.113602},
  url = {https://link.aps.org/doi/10.1103/PhysRevLett.107.113602}
}

@article{HeppLieb1973,
  author = {Hepp, Klaus and Lieb, Elliott H.},
  title = {On the superradiant phase transition for molecules in a quantized radiation field: the Dicke maser model},
  journal = {Annals of Physics},
  volume = {76},
  number = {2},
  pages = {360--404},
  year = {1973},
  doi = {10.1016/0003-4916(73)90039-0}
}

@article{mendoncca2025,
  title = {Role of Matter Interactions in Superradiant Phenomena},
  author = {Mendon\ifmmode \mbox{\c{c}}\else \c{c}\fi{}a, Jo\~ao Pedro and Jachymski, Krzysztof and Wang, Yao},
  journal = {Phys. Rev. Lett.},
  volume = {135},
  issue = {13},
  pages = {133601},
  numpages = {7},
  year = {2025},
  month = {Sep},
  publisher = {American Physical Society},
  doi = {10.1103/z8gv-7yyk},
  url = {https://link.aps.org/doi/10.1103/z8gv-7yyk}
}

@misc{lenk2022collective,
      title={Collective theory for an interacting solid in a single-mode cavity}, 
      author={Katharina Lenk and Jiajun Li and Philipp Werner and Martin Eckstein},
      year={2022},
      eprint={2205.05559},
      archivePrefix={arXiv},
      primaryClass={cond-mat.str-el},
      url={https://arxiv.org/abs/2205.05559}, 
}

@article{roman2025bound,
url = {https://doi.org/10.1515/nanoph-2024-0568},
title = {Bound polariton states in the Dicke–Ising model},
title = {},
author = {Juan Román-Roche and Álvaro Gómez-León and Fernando Luis and David Zueco},
pages = {2053--2064},
volume = {14},
number = {11},
journal = {Nanophotonics},
doi = {doi:10.1515/nanoph-2024-0568},
year = {2025},
lastchecked = {2026-01-11}
}

@article{roman2025linear,
  title = {Linear response theory for cavity QED materials at arbitrary light-matter coupling strengths},
  author = {Rom\'an-Roche, Juan and G\'omez-Le\'on, \'Alvaro and Luis, Fernando and Zueco, David},
  journal = {Phys. Rev. B},
  volume = {111},
  issue = {3},
  pages = {035156},
  numpages = {21},
  year = {2025},
  month = {Jan},
  publisher = {American Physical Society},
  doi = {10.1103/PhysRevB.111.035156},
  url = {https://link.aps.org/doi/10.1103/PhysRevB.111.035156}
}

@article{White1992,
  title = {Density matrix formulation for quantum renormalization groups},
  author = {White, Steven R.},
  journal = {Phys. Rev. Lett.},
  volume = {69},
  issue = {19},
  pages = {2863--2866},
  numpages = {0},
  year = {1992},
  month = {Nov},
  publisher = {American Physical Society},
  doi = {10.1103/PhysRevLett.69.2863},
  url = {https://link.aps.org/doi/10.1103/PhysRevLett.69.2863}
}

@article{White1993,
  author = {White, Steven R.},
  title = {Density-matrix algorithms for quantum renormalization groups},
  journal = {Phys. Rev. B},
  volume = {48},
  pages = {10345--10356},
  year = {1993},
  doi = {10.1103/PhysRevB.48.10345}
}

@article{Schollwock2011,
  author = {Schollw{\"o}ck, Ulrich},
  title = {The density-matrix renormalization group in the age of matrix product states},
  journal = {Annals of Physics},
  volume = {326},
  number = {1},
  pages = {96--192},
  year = {2011},
  doi = {10.1016/j.aop.2010.09.012}
}

@article{Fishman2020ITensor,
  author = {Fishman, Matthew and White, Steven R. and Stoudenmire, E. Miles},
  title = {The ITensor Software Library for Tensor Network Calculations},
  journal = {SciPost Phys. Codebases},
  pages = {004},
  year = {2022},
  doi = {10.21468/SciPostPhysCodeb.4}
}

@article{reslen2004direct,
doi = {10.1209/epl/i2004-10313-4},
url = {https://doi.org/10.1209/epl/i2004-10313-4},
year = {2004},
month = {dec},
publisher = {},
volume = {69},
number = {1},
pages = {8},
author = {J. Reslen and L. Quiroga and N. F. Johnson},
title = {Direct equivalence between quantum phase transition 
phenomena in radiation-matter and magnetic systems: 
Scaling of entanglement},
journal = {Europhysics Letters}
}

@article{anderson1965iterative,
author = {Anderson, Donald G.},
title = {Iterative Procedures for Nonlinear Integral Equations},
year = {1965},
issue_date = {Oct. 1965},
publisher = {Association for Computing Machinery},
address = {New York, NY, USA},
volume = {12},
number = {4},
issn = {0004-5411},
url = {https://doi.org/10.1145/321296.321305},
doi = {10.1145/321296.321305},
journal = {J. ACM},
month = oct,
pages = {547–560},
numpages = {14}
}

@article{ixert2016nonperturbative,
  title = {Nonperturbative linked-cluster expansions in long-range ordered quantum systems},
  author = {Ixert, Dominik and Schmidt, Kai Phillip},
  journal = {Phys. Rev. B},
  volume = {94},
  issue = {19},
  pages = {195133},
  numpages = {16},
  year = {2016},
  month = {Nov},
  publisher = {American Physical Society},
  doi = {10.1103/PhysRevB.94.195133},
  url = {https://link.aps.org/doi/10.1103/PhysRevB.94.195133}
}

@dataset{langhelddata,
  author       = {Langheld, Anja and Hörmann, Maximilian and Schmidt, Kai Phillip},
  title        = {Raw data to ``Quantum phase diagrams of Dicke-Ising models by a wormhole algorithm''},
  year         = {2025},
  publisher    = {Zenodo},
  doi          = {10.5281/zenodo.15774230},
  url          = {https://doi.org/10.5281/zenodo.15774230},
}

@article{baumann2010,
	author = {Baumann, Kristian and Guerlin, Christine and Brennecke, Ferdinand and Esslinger, Tilman},
	date = {2010/04/01},
	doi = {10.1038/nature09009},
	id = {Baumann2010},
	isbn = {1476-4687},
	journal = {Nature},
	number = {7293},
	pages = {1301--1306},
	title = {Dicke quantum phase transition with a superfluid gas in an optical cavity},
	url = {https://doi.org/10.1038/nature09009},
	volume = {464},
	year = {2010}}

@article{wang1973,
  title = {Phase Transition in the Dicke Model of Superradiance},
  author = {Wang, Y. K. and Hioe, F. T.},
  journal = {Phys. Rev. A},
  volume = {7},
  issue = {3},
  pages = {831--836},
  numpages = {0},
  year = {1973},
  month = {Mar},
  publisher = {American Physical Society},
  doi = {10.1103/PhysRevA.7.831},
  url = {https://link.aps.org/doi/10.1103/PhysRevA.7.831}
}

@article{Baumann2011,
  title = {Exploring Symmetry Breaking at the Dicke Quantum Phase Transition},
  author = {Baumann, K. and Mottl, R. and Brennecke, F. and Esslinger, T.},
  journal = {Phys. Rev. Lett.},
  volume = {107},
  issue = {14},
  pages = {140402},
  numpages = {5},
  year = {2011},
  month = {Sep},
  publisher = {American Physical Society},
  doi = {10.1103/PhysRevLett.107.140402},
  url = {https://link.aps.org/doi/10.1103/PhysRevLett.107.140402}
}

@article{fink2009,
  title = {Dressed Collective Qubit States and the Tavis-Cummings Model in Circuit QED},
  author = {Fink, J. M. and Bianchetti, R. and Baur, M. and G\"oppl, M. and Steffen, L. and Filipp, S. and Leek, P. J. and Blais, A. and Wallraff, A.},
  journal = {Phys. Rev. Lett.},
  volume = {103},
  issue = {8},
  pages = {083601},
  numpages = {4},
  year = {2009},
  month = {Aug},
  publisher = {American Physical Society},
  doi = {10.1103/PhysRevLett.103.083601},
  url = {https://link.aps.org/doi/10.1103/PhysRevLett.103.083601}
}

@article{dimer2007,
  title = {Proposed realization of the Dicke-model quantum phase transition in an optical cavity QED system},
  author = {Dimer, F. and Estienne, B. and Parkins, A. S. and Carmichael, H. J.},
  journal = {Phys. Rev. A},
  volume = {75},
  issue = {1},
  pages = {013804},
  numpages = {14},
  year = {2007},
  month = {Jan},
  publisher = {American Physical Society},
  doi = {10.1103/PhysRevA.75.013804},
  url = {https://link.aps.org/doi/10.1103/PhysRevA.75.013804}
}

@article{gammelmark2011,
	author = {Gammelmark, S{\o}ren and M{\o}lmer, Klaus},
	doi = {10.1088/1367-2630/13/5/053035},
	journal = {New Journal of Physics},
	month = {may},
	number = {5},
	pages = {053035},
	title = {Phase transitions and Heisenberg limited metrology in an Ising chain interacting with a single-mode cavity field},
	url = {https://doi.org/10.1088/1367-2630/13/5/053035},
	volume = {13},
	year = {2011}}

@article{lee2004,
  title = {First-Order Superradiant Phase Transitions in a Multiqubit Cavity System},
  author = {Lee, Chiu Fan and Johnson, Neil F.},
  journal = {Phys. Rev. Lett.},
  volume = {93},
  issue = {8},
  pages = {083001},
  numpages = {4},
  year = {2004},
  month = {Aug},
  publisher = {American Physical Society},
  doi = {10.1103/PhysRevLett.93.083001},
  url = {https://link.aps.org/doi/10.1103/PhysRevLett.93.083001}
}

@misc{comment2025,
      title={Comment on "Role of Matter Interactions in Superradiant Phenomena"}, 
      author={Max Hörmann and Anja Langheld and Jonas Leibig and Andreas Schellenberger and Kai Phillip Schmidt},
      year={2025},
      eprint={2511.08452},
      archivePrefix={arXiv},
      primaryClass={quant-ph},
      url={https://arxiv.org/abs/2511.08452}, 
}

@article{rigol2006,
  title = {Numerical Linked-Cluster Approach to Quantum Lattice Models},
  author = {Rigol, Marcos and Bryant, Tyler and Singh, Rajiv R. P.},
  journal = {Phys. Rev. Lett.},
  volume = {97},
  issue = {18},
  pages = {187202},
  numpages = {4},
  year = {2006},
  month = {Nov},
  publisher = {American Physical Society},
  doi = {10.1103/PhysRevLett.97.187202},
  url = {https://link.aps.org/doi/10.1103/PhysRevLett.97.187202}
}

@article{rigol2007a,
  title = {Numerical linked-cluster algorithms. I. Spin systems on square, triangular, and kagom\'e lattices},
  author = {Rigol, Marcos and Bryant, Tyler and Singh, Rajiv R. P.},
  journal = {Phys. Rev. E},
  volume = {75},
  issue = {6},
  pages = {061118},
  numpages = {13},
  year = {2007},
  month = {Jun},
  publisher = {American Physical Society},
  doi = {10.1103/PhysRevE.75.061118},
  url = {https://link.aps.org/doi/10.1103/PhysRevE.75.061118}
}

@article{rigol2007b,
  title = {Numerical linked-cluster algorithms. II. $t\text{\ensuremath{-}}J$ models on the square lattice},
  author = {Rigol, Marcos and Bryant, Tyler and Singh, Rajiv R. P.},
  journal = {Phys. Rev. E},
  volume = {75},
  issue = {6},
  pages = {061119},
  numpages = {6},
  year = {2007},
  month = {Jun},
  publisher = {American Physical Society},
  doi = {10.1103/PhysRevE.75.061119},
  url = {https://link.aps.org/doi/10.1103/PhysRevE.75.061119}
}

@misc{verstraete2004,
      title={Renormalization algorithms for Quantum-Many Body Systems in two and higher dimensions}, 
      author={F. Verstraete and J. I. Cirac},
      year={2004},
      eprint={cond-mat/0407066},
      archivePrefix={arXiv},
      primaryClass={cond-mat.str-el},
      url={https://arxiv.org/abs/cond-mat/0407066}, 
}

@article{stoudemire14,
  title = {Corner contribution to the entanglement entropy of strongly interacting O(2) quantum critical systems in 2+1 dimensions},
  author = {Stoudenmire, E. M. and Gustainis, Peter and Johal, Ravi and Wessel, Stefan and Melko, Roger G.},
  journal = {Phys. Rev. B},
  volume = {90},
  issue = {23},
  pages = {235106},
  numpages = {6},
  year = {2014},
  month = {Dec},
  publisher = {American Physical Society},
  doi = {10.1103/PhysRevB.90.235106},
  url = {https://link.aps.org/doi/10.1103/PhysRevB.90.235106}
}


\end{document}